\documentclass[journal=nalefd,manuscript=communication]{achemso}

\usepackage[version=3]{mhchem} 

\author{Siddharth Tallur}
\affiliation[Cornell University]
{OxideMEMS Lab, Cornell University, Ithaca, NY, USA.}
\author{Sunil A. Bhave}
\email{sunil@ece.cornell.edu}
\affiliation[Cornell University]
{OxideMEMS Lab, Cornell University, Ithaca, NY, USA.}

\title[\texttt{achemso} detector]
{A silicon electromechanical photodetector}

\begin{document}
\begin{abstract}
Opto-mechanical systems have enabled wide-band optical frequency conversion and multi-channel all-optical radio frequency amplification. Realization of an on-chip silicon communication platform is limited by photodetectors needed to convert optical information to electrical signals for further signal processing. In this paper we present a coupled silicon micro-resonator, which converts near-IR optical intensity modulation at 174.2MHz and 1.198GHz into motional electrical current. This device emulates a photodetector which detects intensity modulation of continuous wave laser light in the full-width-at-half-maximum bandwidth of the mechanical resonance. The resonant principle of operation eliminates dark current challenges associated with convetional photodetectors.
\end{abstract}


Cavity opto-mechanical systems offer a unique platform wherein the coherent interaction rate is larger than the thermal decoherence rate of the system, as realized in ground-state cooling experiments \cite{cooling}. This interplay of light and motion opens up an array of novel applications in classical and quantum optics communication\cite{vtt,stamperkurn, darkmode}. In the classical regime, opto-mechanical systems have enabled wide-band optical frequency conversion \cite{painterwavconv} and multi-channel all-optical radio frequency amplification \cite{moli}. Realization of an on-chip silicon communication platform is limited by photo-detectors needed to convert optical information to electrical signals for further signal processing.

Classical and quantum information transfer and storage utilize photons for long range communication \cite{agr, zoller, comb}. Photons are appealing for such applications on account of their weak interaction with the environment and resiliency to thermal noise due to their high frequency. On the other hand, acoustic phonons have lower bandwidth and experience significantly higher losses associated with transmission, but can be delayed and stored for longer times and can interact resonantly with RF-microwave electronic systems.  It has been proposed in the past that a hybrid phononic-photonic system could perform a range of tasks unreachable by purely photonic and phononic systems \cite{trans, kartik, cryst}. Furthermore, such a system should also be capable of being directly integrated with electronics used for processing radio-frequency (RF) signals. Recent research efforts have enabled the realization of opto-mechanical systems where photons are used to manipulate mechanical vibrations and vice versa \cite{carmon, cryst, suresh, bowen, mingwu, zipper}. Achieving this in an all silicon chip-scale platform has been pursued with great zest, as silicon processing offers benefits in terms of lowering manufacturing cost, obtaining high yields, and promises seamless on-chip integration with electronics. However, as the field of silicon photonics has grown, a theme that has emerged is that as a platform, silicon does not provide best-in-class devices for all tasks \cite{myth}. True monolithic integration of photonics devices with cutting-edge 28nm or smaller CMOS processes is a very challenging task. Making process modifications to support such integration will fundamentally change the models for transistors, at the cost of degrading their performance. Not modifying the process is an option, and recently researchers have shown that some photonic functionality can be integrated with minimal post-processing in a silicon-on-insulator (SOI) CMOS process \cite{orcutt}. A key component for CMOS compatible silicon photonics is a photodiode capable of detecting light in the near infrared. Various CMOS compatible photodiodes have been demonstrated \cite{geis, preston, derose, kimerling} but they suffer from complex processing steps to overcome lattice mismatch issues, large area consumption and susceptibility to dark current. Here, we present an alternate photodetector that responds opto-mechanically to optical intensity modulation. The device is comprised of a silicon micro-resonator designed in a CMOS silicon-on-insulator (SOI) platform, which converts optical intensity modulation into motional electrical current, thus converting the signal from photon$\rightarrow$phonon$\rightarrow$electron, instead of depending on an avalanche or photoelectric process in a non-silicon material. This scheme is universal, and could potentially be of interest to opto-mechanical resonators fabricated in piezoelectric materials such as aluminum nitride \cite{tangaln, cleland}, and materials with attractive mechanical and optical properties such as single crystalline diamond \cite{loncar}.

\begin{figure}[htbp]
\centering
\includegraphics[width=8cm]{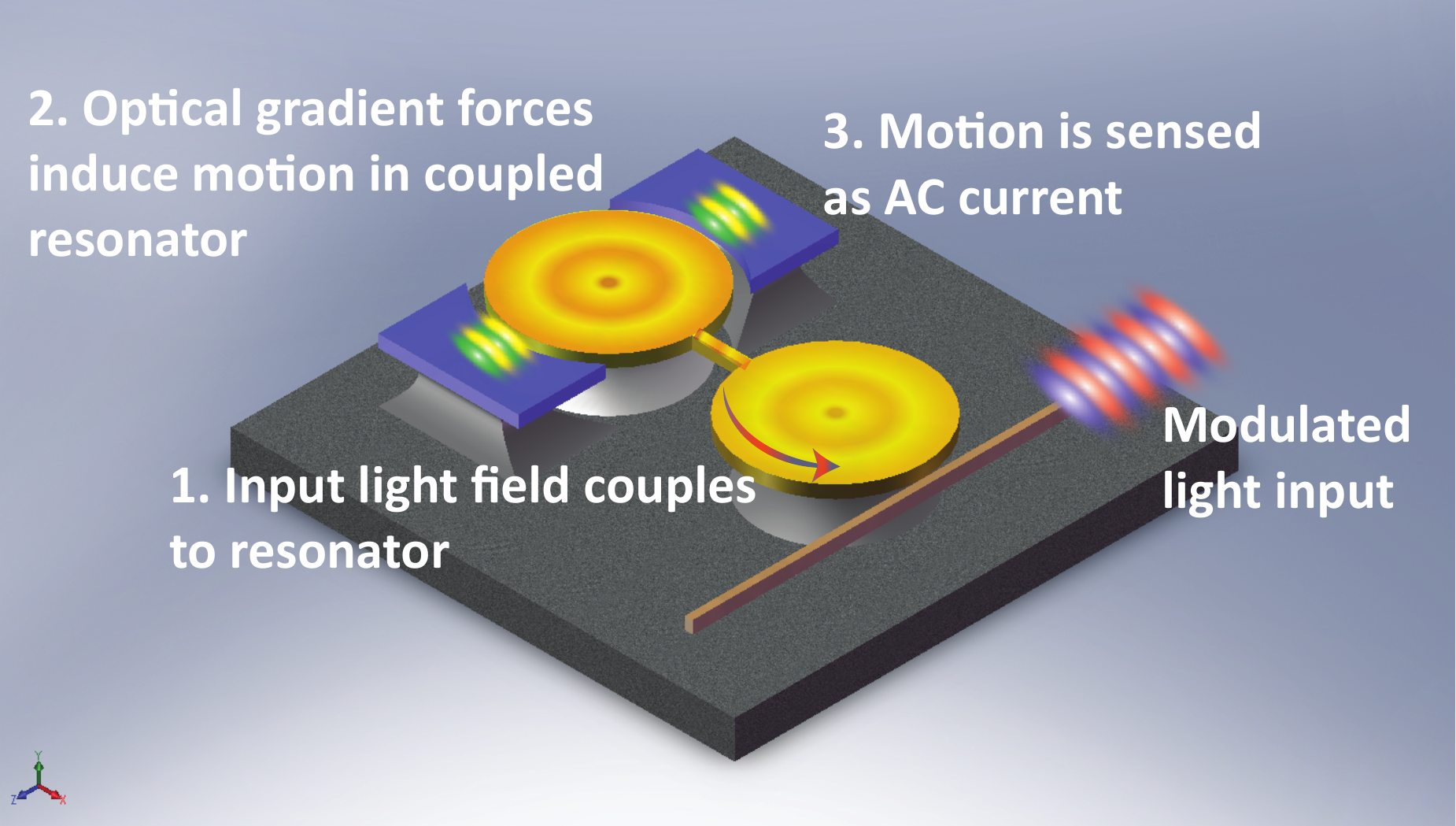}
\caption{Illustration of operation of the electromechanical photodetector.}
\label{fig_illus}
\end{figure}

Figure \ref{fig_illus} shows an illustration showcasing the principle of operation of this detector. The device comprises of two mechanical resonators that are coupled to each other via a mechanical beam. The resonator that is located in close proximity to the waveguide serves as an opto-mechanical resonator and is actuated via optical gradient forces generated by the modulated input light field. These mechanical vibrations $\widetilde{x}(\Omega)$ are coupled to the other mechanical resonator flanked by electrodes via the coupling beam. The motion of this resonator is sensed via a high dynamic range electrostatic capacitive sense scheme, resulting in an AC current $(\widetilde{i}_{out}(\Omega))$ generated on account of modulation of the capacitance formed by the air gap between the resonator and the electrode.

Consider an input pump field $A_p(t)=A_{p0}(t)+\delta$$A_p(t)$, where the first term denotes static field amplitude and the second term is a dynamic modulated term. Correspondingly the intra-cavity dropped field can be expressed similarly as $a_p(t)=a_{p0}(t)+\delta$$a_p(t)$, where:

\begin{equation}
a_{p0} = \frac{j\sqrt{\Gamma_{ex}}}{\frac{\Gamma_{tot}}{2}-j\Delta_p}A_{p0}
\label{eqna}
\end{equation}

Here $\Delta_p$ is the detuning of the laser $(\omega_p)$ from the cavity resonance $(\omega_0)$ and $\Gamma_{in}$, $\Gamma_{ex}$ and $\Gamma_{tot}$ are the intrinsic cavity photon decay rate, photon decay rate associated with coupling to the cavity, and photon decay rate of the loaded optical cavity respectively. The intra-cavity field is normalized to the intra-cavity energy, $U_{p}=\left|a_{p}\right|^2$. The optical gradient force acting on the opto-mechanical resonator can then be expressed as follows \cite{jesse}:

\begin{equation}
F_{grad} = -\frac{g_{OM}U_p}{\omega_p}
\end{equation}

$g_{OM}$ denotes the opto-mechanical coupling coefficient. The gradient force consists of two terms: $F_{grad}(t)=F_{grad,0}(t)+\delta$$F_{grad}(t)$. The first term is a static force whereas the second term is the dynamic component related to the laser light modulation $\delta$$U_p(t)$, given by:

\begin{equation}
\delta F_{grad}(t) = -\frac{g_{OM}\delta U_p}{\omega_p} = -\frac{g_{OM}}{\omega_p}\left[a_{p0}^*\delta a_p(t)+a_{p0}\delta a_p^*(t)\right]
\end{equation}

The mechanical motion $(x)$ of the cavity follows: $\ddot{x}+\Gamma_m\dot{x}+\Omega_m^2x=\frac{F_{grad,0}(t)+\delta F_{grad}(t)+F_T}{m_{eff}}$, where $m_{eff}$ is the effective mass of the mechanical mode with frequency $\Omega_m$, $\Gamma_m$ is the intrinsic mechanical damping rate, and $F_T$ is the thermal Langevin force responsible for the thermal Brownian motion. The dynamic displacement of the resonator is affected largely by the dynamic gradient force, as the thermal Langevin force is relatively much smaller in magnitude. The back-action of the mechanical motion changes the value of the resonant frequency and damping rate of mechanical motion. The spectral response of this force is given by \cite{jesse}:

\begin{equation}
f_{0}(\Omega) = -\frac{2g_{OM}^2 U_{p0} \Delta_p}{\omega_p} \frac{\Delta_p^2-\Omega^2+\left(\frac{\Gamma_{tot}}{2}\right)^2+j\Gamma_{tot}\Omega}{\left[(\Delta_p+\Omega)^2+\left(\frac{\Gamma_{tot}}{2}\right)^2\right]\left[(\Delta_p-\Omega)^2+\left(\frac{\Gamma_{tot}}{2}\right)^2\right]}
\end{equation}

Define $\ell(\Omega)=\Omega_m^2-\Omega^2-j\Gamma_m\Omega-\frac{f_o(\Omega)}{m_{eff}}$. The dynamic mechanical displacement on account of the optical gradient force is given by:

\begin{equation}
\widetilde{x}(\Omega) =  \frac{j\sqrt{\Gamma_{ex}}g_{OM}}{m_{eff}\omega_p\ell(\Omega)}\left[\frac{a_{p0}^*\delta \widetilde{A}_p(\Omega)}{j(\Delta_p+\Omega)-\frac{\Gamma_{tot}}{2}}+\frac{a_{p0}\delta \widetilde{A}_p^*(-\Omega)}{j(\Delta_p-\Omega)+\frac{\Gamma_{tot}}{2}}\right]
\label{eqnx}
\end{equation}

Here $\delta \widetilde{A}_p$ is the Fourier domain representation of the modulated input light field. This expression is complete in the sense that it accounts for motion actuated due to the optical gradient force acting on the resonator, and also the back-action induced by the motion on the optical field. Substituting equation \ref{eqna} into equation \ref{eqnx} yields:

\begin{equation}
\widetilde{x}(\Omega) =  \frac{-\Gamma_{ex}g_{OM}}{m_{eff}\omega_p\ell(\Omega)\left(\frac{\Gamma_{tot}}{2}-j\Delta_p\right)}\left[\frac{A_{p0}^*\delta \widetilde{A}_p(\Omega)}{j(\Delta_p+\Omega)-\frac{\Gamma_{tot}}{2}}+\frac{A_{p0}\delta \widetilde{A}_p^*(-\Omega)}{j(\Delta_p-\Omega)+\frac{\Gamma_{tot}}{2}}\right]
\label{eqnax}
\end{equation}

In the unresolved sideband regime, the equation above reduces to the limit $\displaystyle\lim_{\Omega \to 0}\widetilde{x}(\Omega)$:

\begin{align}
\widetilde{x}(\Omega) &= \frac{-\Gamma_{ex}g_{OM}\left[-j\Delta_p\left(A_{p0}^*\delta \widetilde{A}_p(\Omega)+A_{p0}\delta \widetilde{A}_p^*(-\Omega)\right)-\frac{\Gamma_{tot}}{2}\left(A_{p0}^*\delta \widetilde{A}_p(\Omega)-A_{p0}\delta \widetilde{A}_p^*(-\Omega)\right)\right]}{m_{eff}\omega_p\ell(\Omega)\left(\frac{\Gamma_{tot}}{2}-j\Delta_p\right)\left[\Delta_p^2+\frac{\Gamma_{tot}^2}{4}\right]} \\
&\approx \frac{-\Gamma_{ex}g_{OM}\left[-j\Delta_p\left(A_{p0}^*\delta \widetilde{A}_p(\Omega)+A_{p0}\delta \widetilde{A}_p^*(-\Omega)\right)\right]}{m_{eff}\omega_p\ell(\Omega)\left(\frac{\Gamma_{tot}}{2}-j\Delta_p\right)\left[\Delta_p^2+\frac{\Gamma_{tot}^2}{4}\right]} \\
&= \frac{j\Delta _p \Gamma_{ex}g_{OM}\delta P_{in}(\Omega)}{m_{eff}\omega_p\ell(\Omega)\left(\frac{\Gamma_{tot}}{2}-j\Delta_p\right)\left[\Delta_p^2+\frac{\Gamma_{tot}^2}{4}\right]}
\label{eqnax_ressb}
\end{align}

The derivation above assumes that the displacement amplitude of the resonator is small i.e. the perturbation of the detuning on account of mechanical motion is very small compared to the unperturbed detuning $\left(\frac{x(t)}{R}\omega_0\ll\Delta_p\right)$. In the large amplitude regime, where the detuning oscillates between large positive and negative values, the small signal model derived above fails to hold. An extensive study of opto-mechanical oscillation amplitudes was recently presented by Poot et al. \cite{Poot}. The optical backaction on the resonator enables radiation pressure induced self-sustained oscillations whose limit cycle is set by the dynamic range of the cavity. This sets a maximum limit on the amplitude of mechanical motion \cite{Poot}, which would amount to saturation of the motional current generated by the detector.

If $C$ denotes the capacitance formed by the air gap between the resonator and the electrode, and $g$ denotes the resonator-electrode gap, the motional current flowing into the electrode in response to the motion of the resonator and applied DC voltage, $V_{dc}$ is expressed as:

\begin{equation}
i_{out}(t) = V_{dc}\frac{dC}{dt} = V_{dc}\frac{dC}{dg}\frac{dg}{dt}
\end{equation}

the dynamic component of this current can then be written down as:

\begin{equation}
\widetilde{i}_{out}(\Omega) = V_{dc}\frac{dC}{dg}j\Omega\widetilde{x}(\Omega) = V_{dc}\frac{\epsilon_0Rh\theta}{g^2}j\Omega\widetilde{x}(\Omega)
\label{eqni}
\end{equation}

Here $R$, $h$ and $\theta$ denote the outer radius of the resonator, device thickness and the electrode-resonator overlap angle respectively. Substituting equation \ref{eqnax_ressb} into equation \ref{eqni}, we get:

\begin{equation}
\frac{\widetilde{i}_{out}(\Omega)}{\delta P_{in}(\Omega)} =\frac{-\Delta _p \Gamma_{ex}g_{OM}V_{dc}\epsilon_0Rh\theta\Omega}{g^2m_{eff}\omega_p\ell(\Omega)\left(\frac{\Gamma_{tot}}{2}-j\Delta_p\right)\left[\Delta_p^2+\frac{\Gamma_{tot}^2}{4}\right]}
\label{eqniP}
\end{equation}

The expression derived above in equation \ref{eqniP} can be interpreted as the ``electromechanical responsivity'' $\left(\Re_{em}\right)$ of the detector.

\begin{figure}[htbp]
\centering
\includegraphics[width=8cm]{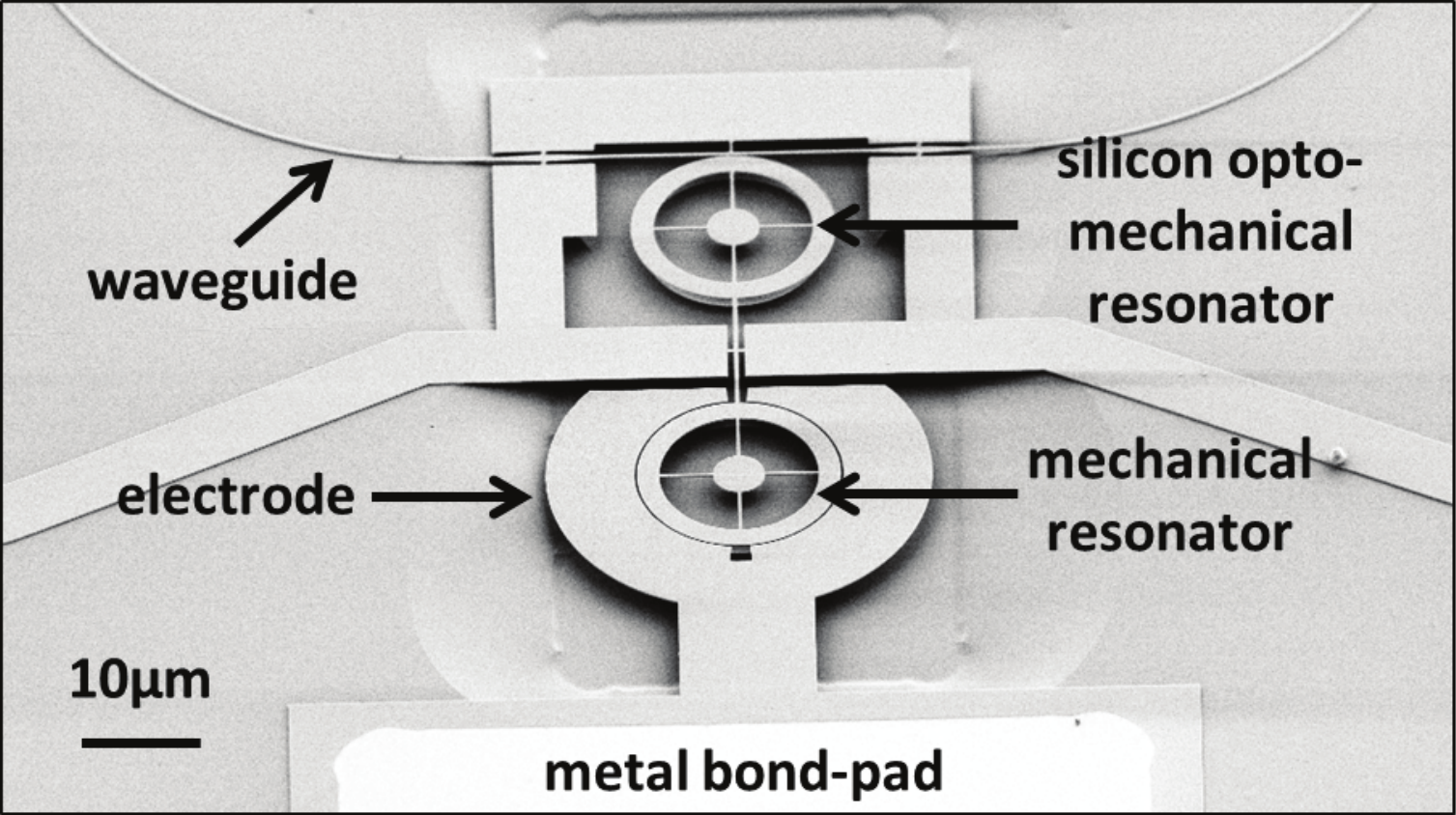}
\caption{Scanning electron micrograph of the silicon electro-mechanical detector integrated with electrodes and a waveguide. The lithographically defined gap between the electrode and resonator is 50nm.}
\label{fig_sem}
\end{figure}

We choose a coupled micro-ring geometry \cite{suresh} for the photodetector. The motional current amplitude varies as inverse-square of the resonator-electrode gap $(g)$ as shown in equation \ref{eqniP}, and hence it is desirable to reduce the gap to boost the detection efficiency. The smallest gaps that we could define were limited by the resolution of lithography and hence the gap is set to 50nm in our device to enable efficient electrostatic transduction. The outer radius $(R)$ of the ring resonators is 9.5$\mu$m and the silicon device layer thickness $(h)$ is 220nm. As evident from equation \ref{eqniP}, the greater the electrode overlap angle $(\theta)$, higher the current measured. The coupling spring forces us to leave a 10$^\circ$ opening in the electrode where it connects to the sense resonator. To maintain symmetry of design and ensure mechanical momentum balance we introduce another 10$^\circ$ opening on the diametrically opposite section of the electrode, as seen in the Scanning Electron micrograph (SEM) in Figure \ref{fig_sem}.

Fabricating the photodetector involves a four mask process flow on a custom silicon-on-insulator (SOI) wafer (undoped 250 nm device layer for low optical loss and 3$\mu$m thick buried oxide for isolation of the waveguides on device layer from the silicon substrate). The top silicon is thermally oxidized to obtain a thin oxide hard mask layer of thickness 60nm atop a 220nm thick silicon device layer. ma-N 2403 electron beam resist is spun on top of the oxide and patterned using electron beam lithography. The patterns are transferred into the silicon dioxide using a CHF$_3$/O$_2$ based reactive ion etcher and then into the silicon device layer using a chlorine based reactive ion etch. A layer of SPR-220 3.0 photoresist is spun and a second mask is used to pattern windows above the mechanical resonator, the electrical routing beams and the bond-pads. This is followed by a boron ion implantation and nitrogen annealing to reduce the resistivity of these structures. A third mask is then used to deposit metal over the bond pads for improved electrical contact. A layer of LOR-5A followed by a layer of SPR-220 3.0 is spun and the bond pads are exposed via contact alignment photolithography. This is followed by evaporation of 25nm nickel on the sample. Nickel forms a good ohmic contact with silicon, and is hence chosen as the bottommost metal. After evaporating nickel, we evaporate 25nm titanium and 50nm platinum. Platinum is used as the top metal as it makes good electrical contact with the Cascade Air Coplanar Probe (ACP) RF probe used to interrogate these devices. However platinum does not adhere well to nickel, and hence titanium is used as an adhesion layer. The photoresist is dissolved in Microposit remover solvent 1165 to leave metal only atop bond-pads. A fourth mask is used to pattern release windows near the resonator using SPR-220 3.0 photoresist, followed by a timed release etch in buffered oxide etchant to undercut the devices. The samples are then dried using a critical point dryer to prevent stiction.

\begin{figure}[htbp]
\centering
\includegraphics[width=8cm]{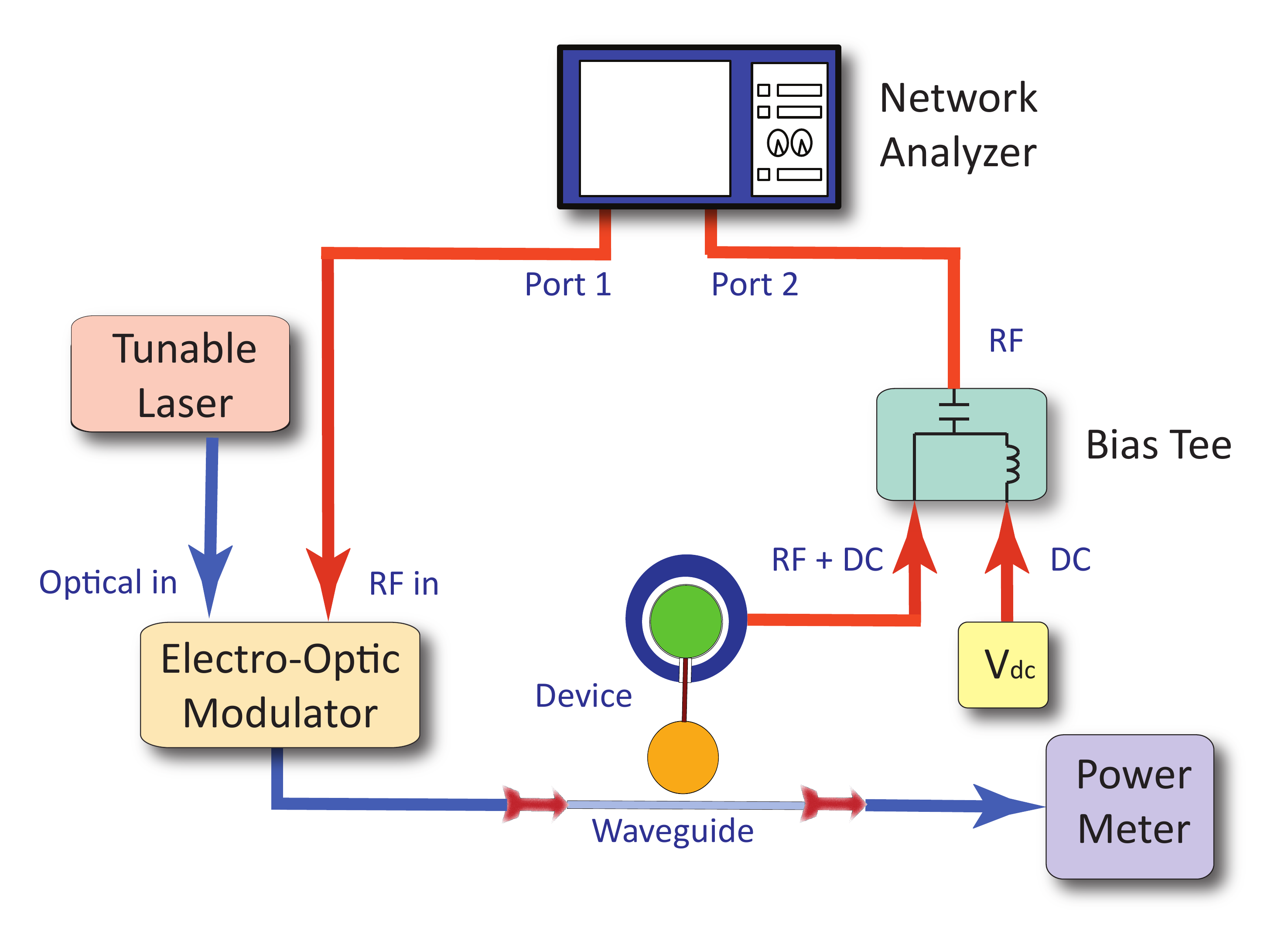}
\caption{Schematic of the experimental setup used to characterize the silicon detector. Mechanical motion is actuated by modulating continuous wave laser light coupled into the opto-mechanical resonator using a Photline MXAN-LN-10 electro-optic modulator. The mechanical motion is sensed via electrostatic capacitive actuation by applying V$_{dc}$ = 40V DC voltage at the electrode using a bias tee. A network analyzer is used to measure the 2-port transmission of the device.}
\label{fig_setup}
\end{figure}

Figure \ref{fig_setup} shows a schematic of the experimental setup used to characterize the photodetector performance. Light from a Santec TSL-510 tunable diode laser is modulated with a Photline MXAN-LN-10 lithium niobate electro-optic intensity modulator (EOM). An Agilent N5230A network analyzer is used to characterize the detector efficiency. The input laser light is modulated by connecting the RF input of the modulator to port 1 of the network analyzer. The output power is sensed by connecting the signal from the metal bond pad to port 2 of the network analyzer.

\begin{figure}[ht]
\centering
\includegraphics[width=15cm]{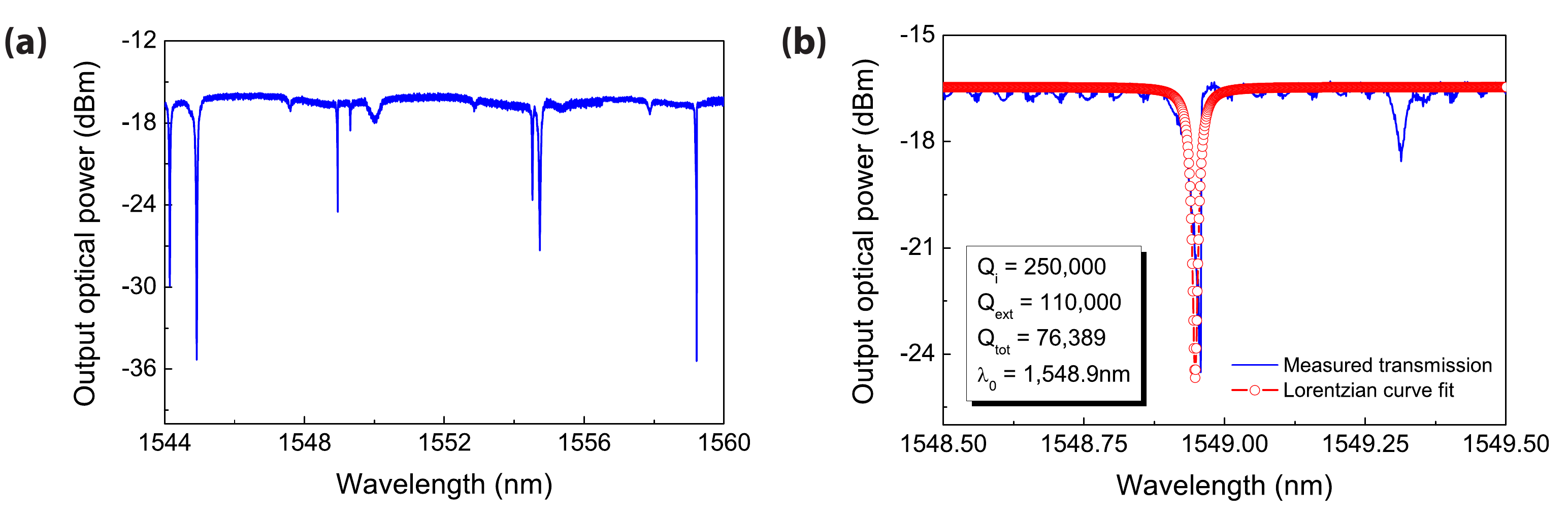}
\caption{(a) Optical spectrum for the opto-mechanical resonator based silicon photodetector. The input laser power is +2dBm. The connectors and grating couplers add 8dB loss at each facet. (b) High optical Q resonance used to operate the photodetector. We intentionally choose an overcoupled resonance in order to measure larger motional current.}
\label{fig_opt}
\end{figure}

The silicon opto-mechanical resonator has many optical resonances in the C-band as seen in Figure \ref{fig_opt}(a). For the purpose of this experiment we choose an overcoupled resonance at 1,548.9nm, with an extinction of 8dB shown in Figure \ref{fig_opt}(b). As derived in equation \ref{eqniP}, the motional current amplitude is proportional to the cavity coupling rate, $\Gamma_{ex}$, and hence operating with an overcoupled resonance is desirable. However, this also reduces the loaded optical quality factor, and hence there is a trade-off associated with overcoupling to the resonator. The rich optical spectrum of the resonator offers us a wide choice of optical resonances to choose from.

\begin{figure}[htbp]
\centering
\includegraphics[width=15cm]{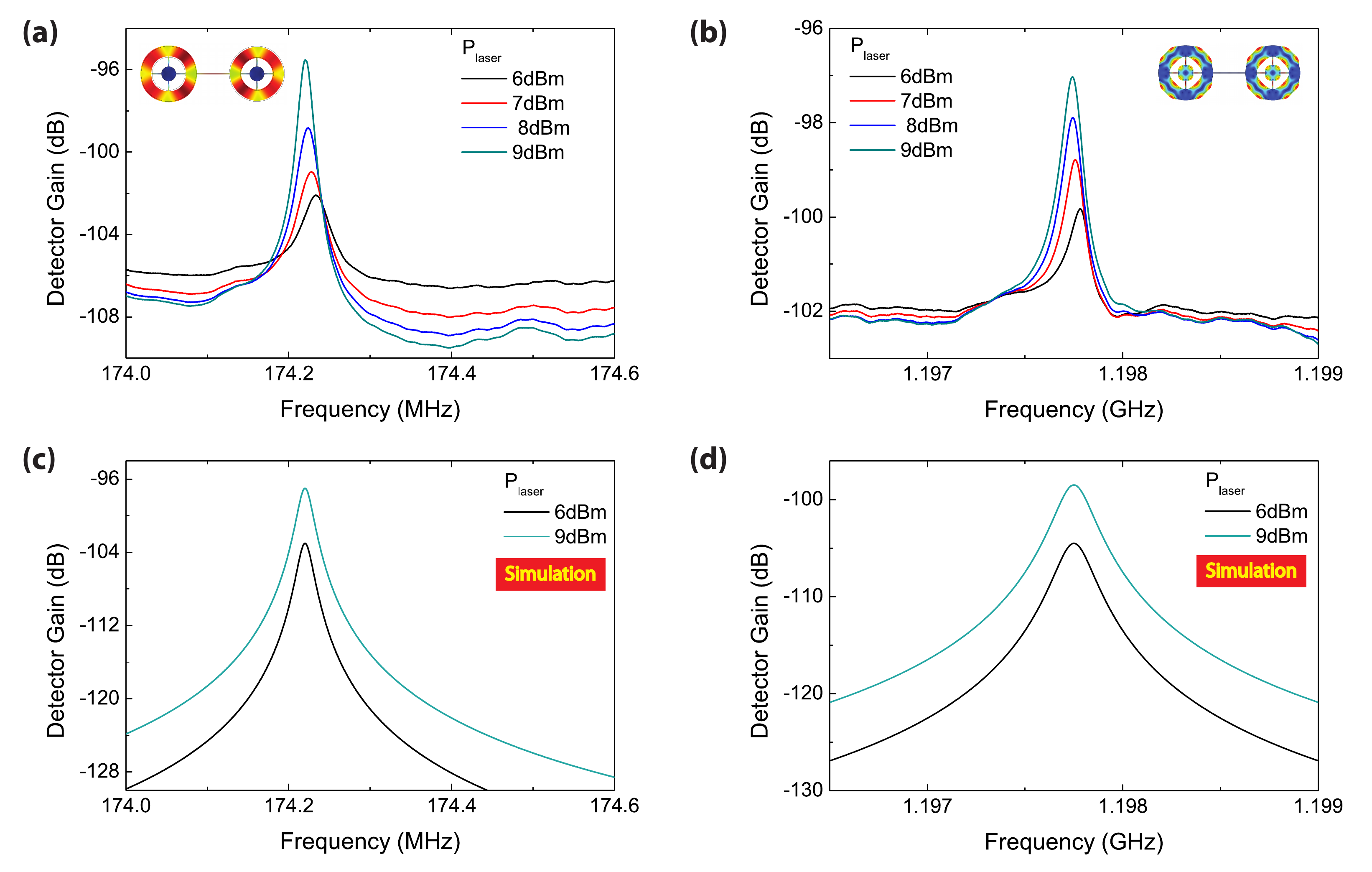}
\caption{Frequency spectra for the detector gain $(P_{out}/P_{in,RF})$ measured using network analyzer. Measured detection of optical modulation for (a) fundamental radial expansion mode at 174.2MHz, and (b) compound radial expansion mode at 1.198GHz. The insets in panels (a) and (b) show the corresponding finite element method (FEM) simulated mechanical mode-shapes. (c) Simulated transmission at frequencies near the fundamental radial expansion resonance frequency for laser power = +6dBm and +9dBm. (d) Simulated transmission at frequencies near the compound radial expansion resonance frequency for laser power = +6dBm and +9dBm. Simulations for expected gain are based on equation \ref{eqni}. Measurements were carried out at room temperature and a pressure of 5mTorr.}
\label{fig_PNA}
\end{figure}

A DC bias voltage of 40V is applied at the metal bond pad using a bias tee. We apply an input RF power $(P_{in,RF})$ of 0dBm at port 1 of the network analyzer, and measure the output RF power $(P_{out})$ at port 2. The transmission of the device operated in this configuration corresponds to the ``gain'' of the photodetector $(P_{out}/P_{in,RF})$. Figure \ref{fig_PNA} shows the measured gain for the detector at various input laser power levels measured at the resonator (by discounting the coupling loss). The signals measured correspond to mechanical vibrations of the fundamental radial expansion mode at 174.2MHz and compound radial expansion mode at 1.198GHz (panel (a) and (b) respectively). The measured gain depends on the input laser power, akin to nanomechanical resonator based microwave amplification reported by Massel et al. \cite{vtt}.

The maximum laser power is set 4dB below the threshold for onset of radiation pressure induced self-oscillations of the fundamental radial expansion mode. Measurements were carried out at room temperature and 5mTorr pressure. The minimum detectable signal at the sense port of the network analyzer (port 2) is set by the receiver's noise floor at this port, which depends on the averaging factor used while carrying out the measurement. An averaging factor of 16 was used in all the measurements to optimize the sensitivity of the network analyzer. The RF power applied at port 1 of the network analyzer is 0dBm. The input laser wavelength is set to the 3-dB off resonance wavelength, and the laser is blue detuned with respect to the optical cavity. Measurements were carried out at room temperature and a pressure of 5mTorr. The measured mechanical quality factors at 174.2MHz and 1.198GHz are 8,700 and 6,300 respectively. Substituting all the experimental parameters into equation \ref{eqni} and calculating the output power, $P_{out} =\frac{\left|i_{out}\right|^2R}{2}$, where $R=50\Omega$ is the load resistance, yields a conversion gain of -97dB at 174.2MHz, and -98dB at 1.198GHz for +9dBm input laser power. The simulated gain at frequencies near 174.2MHz and 1.198GHz are shown in Figure \ref{fig_PNA}(c) and (d), which closely match measured gain values. The shift of the mechanical resonance frequency with increasing input laser power is attributed to thermal heating of the device due to absorption of light coupled into the optical cavity. Native single crystal silicon resonators have negative temperature coefficient of frequency (TCF), and hence larger optical power coupling into the cavity lowers the mechanical resonance frequency. This shift in frequency is negligibly small compared to the intrinsic mechanical resonance frequency $\left(\frac{\Delta \Omega_m}{\Omega_m} \sim 0.1\%\right)$ and hence this effect is not taken into account in simulation.

The conversion of signal from photons to phonons results in a conversion loss of $\Omega_m/\omega_p$, which is to be expected, as evident in equation \ref{eqniP}. The loss values at 174.2MHz and 1.198GHz are -60dB and -52dB respectively. Gaining insights from equation \ref{eqniP}, one could envisage a detector design with larger gain that benefits from higher optical quality factor $\left(Q_{tot}\right)$ resonances, smaller resonator-electrode gaps $\left(g\right)$, and smaller detuning $\left(\Delta_p\right)$. However, choosing a smaller detuning value could potentially launch the device into radiation pressure induced self-oscillations \cite{Poot}, which leads to amplitude saturation.

In conclusion, we have demonstrated an on-chip electro-mechanical detector fabricated on a CMOS SOI platform. The electro-mechanical sense scheme constitutes a high dynamic range detection medium, and we observe efficient signal detection at 174.2MHz all the way up to 1.198GHz. The mechanical resonance frequencies of this device are lithographically defined. The resonant nature of this scheme makes this device a narrow-band detector, whose bandwidth is limited by the quality factor of the mechanical resonance. The sense scheme is universal, and can also be used for detection of optical modulation induced by radiation pressure vibrations, which has successfully been demonstrated at GHz rates in silicon \cite{cryst}. This electro-mechanical detector thus introduces a valuable component in the library of existing novel opto-mechanical devices. To the best of our knowledge, this constitutes the first experimental demonstration of a photon-to-phonon translator.

\section{Author Information}

\textbf{Corresponding Author}\\
*E-mail: sunil@ece.cornell.edu\\
\textbf{Notes}\\
The authors declare no competing financial interests.

\acknowledgement

The authors would like to acknowledge DARPA/MTO's ORCHID program for research support.

%


\end{document}